# LaP$_2$: isostructural to MgB$_2$ with charming superconductivity

Xing Li[1†], Xiaohua Zhang[1,2†], Yong Liu[1], and Guochun Yang[1,2,*]

[1]*State Key Laboratory of Metastable Materials Science & Technology and Key Laboratory for Microstructural Material Physics of Hebei Province, School of Science, Yanshan University, Qinhuangdao 066004, China*
[2]*Centre for Advanced Optoelectronic Functional Materials Research and Key Laboratory for UV Light-Emitting Materials and Technology of Northeast Normal University, Changchun 130024, China*

The exploration of superconductivity dominated by structural units is of great interest in condense matter physics. MgB$_2$, consisting of graphene-like B, becomes a typical representative of traditional superconductors. Phosphorus demonstrates diverse non-planar motifs through sp$^3$ hybridization in allotropes and phosphides. Here, we report that a pressure-stabilized LaP$_2$, isostructural to MgB$_2$, shows superconductivity with a predicted $T_c$ of 22.2 K, which is the highest among already known transition metal phosphides. Besides electron-phonon coupling of graphene-like P, alike the role of B layer in MgB$_2$, La 5d/4f electrons are also responsible for the superconducting transition. Its dynamically stabilized pressure reaches as low as 7 GPa, a desirable feature of pressure-induced superconductors. The distinct P atomic arrangement is attributed to its sp$^2$ hybridization and out-of-plane symmetric distribution of lone pair electrons. Although P is isoelectronic to N and As, we hereby find the different stable stoichiometries, structures, and electronic properties of La phosphides compared with La nitrides/arsenides at high pressure.

## I. INTRODUCTION

Superconductivity refers to a physical phenomenon of materials in which electrical resistance disappears and magnetic flux fields are expelled from them below a critical temperature ($T_c$) [1]. These characters make superconductors become indispensable for the development of high technology. However, their low $T_c$ values have long been a huge obstacle to large-scale and low-cost applications. Recently, the breakthrough finding of near room-temperature superconductivity is promoted by theoretical calculations such as CaH$_6$ [2], LaH$_{10}$ [3], and HfH$_{10}$ [4]. Some of them have been experimentally confirmed [5,6]. On the other hand, this kind of materials exhibits superconductivity dominated by structural units (e.g., sodalite H cage in LaH$_{10}$ and pentagraphenelike H sheet in HfH$_{10}$), providing a clear guidance for the design of high-$T_c$ superconductors. Unfortunately, achieving their superconductivity needs more than one million times the atmospheric pressure [7-9]. Therefore, the preparation of high-$T_c$ superconductors at lower pressure or ambient pressure is of fundamental importance [10,11].

The graphene-like motif is of great interest due to the planar atomic arrangement, sp$^2$ hybridized feature, and intriguing properties [12-14]. For instance, graphene, the first two-dimensional material, shows many new attributes that are not observed in bulk materials, and is regarded as a revolutionary material in the 21st century [15,16]. More interestingly, magic angle graphene, obtained by stacking two graphene sheets with very small twist angles, demonstrates superconductivity and some novel physical phenomena, which has pushed the research of condensed matter physics to a new level [13,17]. MgB$_2$, consisting of a graphene-like B layer [14], becomes a milestone of Bardeen-Cooper-Schrieffer (BCS) superconductors at ambient pressure [18], whose superconductivity associates with the B layer [19,20].

Phosphorus (P) is a magic element mainly due to its diverse motifs (e.g. dimers, chains, rings, tubes, and bucked layers) in elemental solids and phosphides [21,22], which can induce exceptional properties including anisotropy, high selectivity, and high thermal conductivity [23-25]. On the other hand, P has the feature of preferentially forming sp$^3$ hybridization in its motifs, accompanying the repulsive interaction between bonded electrons and lone-pair ones [26]. Thus, non-planar configurations are their common feature.

Pressure can fundamentally modify the bonding pattern. A planar triangular N configuration with a sp$^2$ hybridization has been observed in pressure-induced HeN$_{10}$ [27]. On the other hand, P is the next group-VA element isoelectronic to N. A graphene-like P has been predicted in pressure-induced

P-H framework [28], stabilized by symmetric hydrogen bonding. Thus, there is expectance to obtain planar P configuration by regulating its bonding pattern and valence electron distribution with pressure.

La atom has shown an ability of stabilizing interesting motifs in compounds, such as $LaB_8$ with a $B_{26}$ cage constructed from 12 twisted rhombuses and 6 twisted hexagons, and possessing the largest atom number in the known B cages [29], $LaB_3C_3$ consisting of a truncated octahedral $sp^3$-bonded B-C cage and having a lower synthesis pressure than $SrB_3C_3$ [30], and $LaP_5$ containing a puckered P layer composed of larger $P_{12}$ rings [31], compared with the $P_6$ ring in phosphene [32].

Having this in mind, and considering that phosphides also exhibit superconductivity related to P configurations (e.g., 2.6 K for $Nb_2P_5$ with zigzag P chains [33], 22 K for $KP_2$ with puckered phosphorus layers [34], and 10.2 K for $MgP_2$ with P three-dimensional framework [35]). We can't help but wonder that whether La atoms can regulate the P motif to a graphene-like one under pressure and show a higher $T_c$ value in the corresponding compound. As expected, through structure searches, graphene-like P is stabilized in $LaP_2$, which is isostructural to $MgB_2$. The predicted $T_c$ of $LaP_2$ reaches 22.2 K at 11 GPa, becoming the highest among the reported transition metal (TM) phosphides. Although N and As are isoelectronic to P, their motifs in La-N and La-As compounds are distinct from P configurations.

## II. COMPUTATIONAL DETAILS

First-principles structure search technology plays a leading role in accelerating the discovery of high-$T_c$ superconductors [7-9]. Here, thermodynamically stable candidates of La-N/P/As systems with various chemical compositions at the selected pressures of 1 atm and 25, 50, 100, 200, and 300 GPa are explored by employing the intelligence-based particle-swarm optimization algorithm [36]. The projector augmented wave (PAW) method [37] is adopted with $2s^22p^3$, $3s^23p^3$, $4s^24p^3$ and $5s^25p^65d^16s^2$ as valence electrons for N, P, As, and La atom, respectively. Exchange-correlation potentials are treated with the Perdew-Burke-Ernzerhof (PBE) generalized gradient approximation [38], as implemented in the Vienna Ab initio Simulation Package (VASP) [39]. The cutoff energy of 800 eV and the Monkhorst-Pack scheme with a k-point grid of $2\pi \times 0.03$ Å$^{-1}$ in the Brillouin zone are tested, which can meet the enthalpy convergence of less than 1 meV/atom. Phonon dispersion and electron-phonon coupling (EPC) calculations are performed within the density functional perturbation theory (DFPT) [40], as implemented in the PHONOPY [41] and QUANTUM ESPRESSO [42] packages.

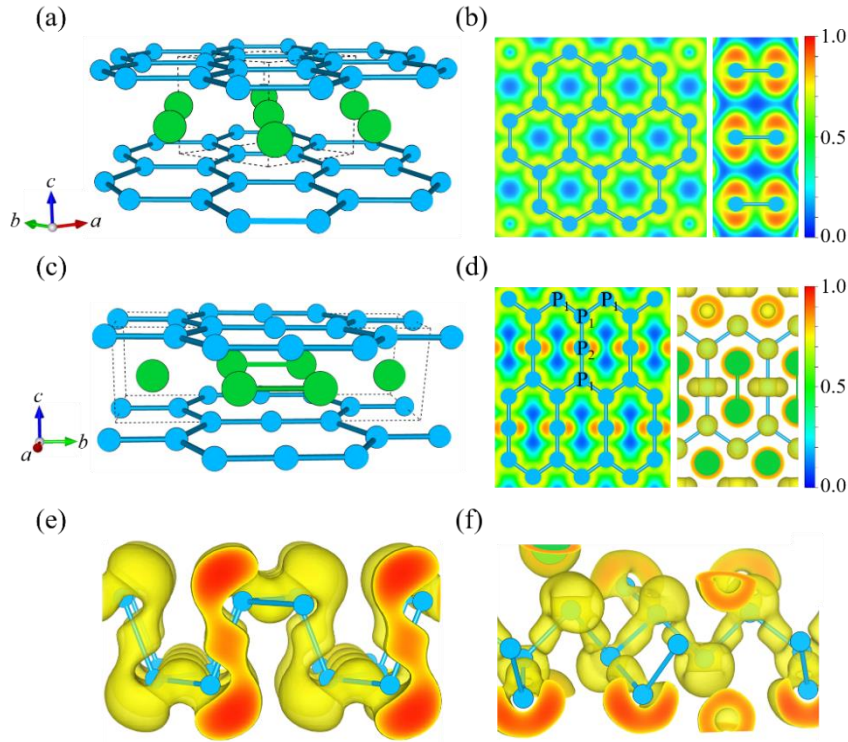

**Fig. 1**. (a) The crystal structure and (b) ELF maps of $P6/mmm$ $LaP_2$ at 16 GPa. (c) The crystal structure and (d) ELF maps of $Cmmm$ $La_2P_3$ at 50 GPa. The green and blue spheres represent La and P atoms, respectively. The ELF isosurfaces with the value of 0.75 of (e) black phosphorus and (f) P framework in $LaP_5$ at ambient pressure.

## III. RESULTS AND DISCUSSION

To build the high-pressure phase diagram of La-P system, we perform extensive structure searches on the stoichiometries of $La_xP_y$ ($x = 1$, $y = 0.5, 1, 1.5, 2, 3\text{-}8$; $x = 3$, $y = 4$) with up to four formula units per simulation cell (Fig. S1b). The energetic stabilities of the $La_xP_y$ compositions are evaluated based on their formation enthalpies relative to the phases of La [43,44] and P [45] at different pressures. In addition to readily reproducing the already known NaCl-type LaP [46], $Cc$ $LaP_2$ [47], $P2_1/m$ $LaP_5$ [31], and $P2_1/n$ $LaP_7$ [48] at 1 atm and the theoretically proposed high-pressure $Imma$ LaP phase [49], we identify several thermodynamically stable chemical compositions (e.g., $La_2P$, LaP, $La_2P_3$, and $LaP_2$) at high pressures. Under compression, these compounds undergo complex structural phase transition. More information about their stable pressure ranges is shown in Fig. S1b. The lack of imaginary frequencies in the phonon dispersion curves confirms their dynamical stability (Fig. S2).

For the stable $La_2P$, LaP, $La_2P_3$, and $LaP_2$ compounds, they demonstrate the feature of P motifs related to P content in compounds, such as isolated P atom, P dimer, P chain, planar P layer, and vertice-sharing P tetrahedra (Fig. S3). Here, we focus on $P6/mmm$ $LaP_2$ and $Cmmm$ $La_2P_3$ consisting of the long-desirable planar P motifs (Fig. 1). The rest of the structures and electronic properties are displayed in Figs. S4-5.

$P6/mmm$ $LaP_2$ is isostructural to $MgB_2$ [14], in which the P atoms form a graphene-like layer in the $ab$ plane, and each La atom is 12-fold coordinated with P atoms in the two adjacent P layers (Fig. 1a). The P-P bond length of 2.32 Å at 16 GPa is slightly larger than 2.20 Å in black phosphorus [50], but it still shows a covalent feature, as confirmed by the electron location function (ELF) analysis (Fig. 1b).

$La_2P_3$ stabilizes into an orthorhombic structure with $Cmmm$ symmetry (Fig. 1c), consisting of one equivalent La atom sitting at 4i position and two inequivalent P atoms occupying 4j and 2c sites, labeled as $P_1$ and $P_2$ (Fig. 1d), respectively. Its striking structural feature is that the two kinds of P atoms form a planar layer via edge-sharing $P_8$ rings in $ab$ plane. The coordination numbers of $P_1$ and $P_2$ atoms are three and two, having a $P_1$-$P_2$ bonding length of 2.33 Å and a $P_1$-$P_1$ of 2.19 Å with covalent feature (Fig. 1d). La atoms adopt a dumbbell-shaped $La_2$ unit, and each La atom is 10-fold coordinated with P atoms.

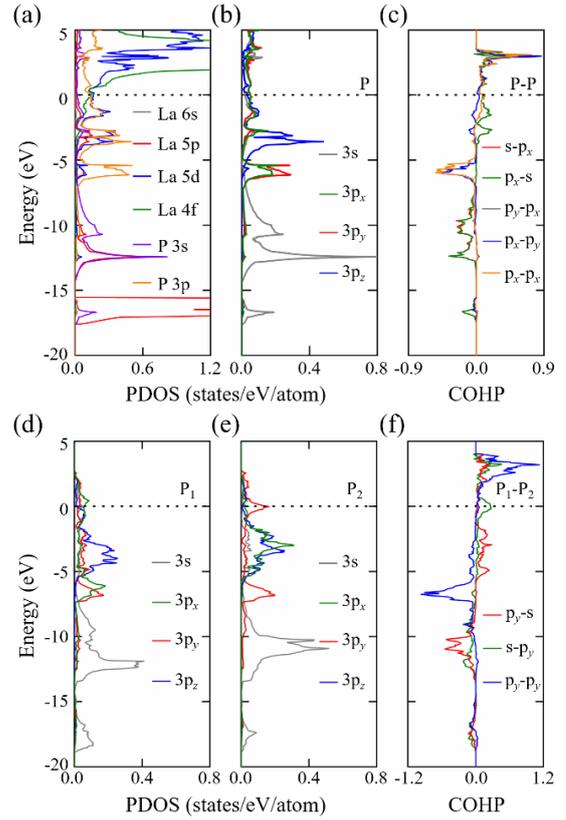

**Fig. 2**. (a) The PDOS of La and P atoms, (b) the orbital-resolved PDOS of P atom, and (c) the COHP of P-P interaction in $P6/mmm$ $LaP_2$ at 16 GPa. The orbital-resolved PDOS of (d) $P_1$ atom and (e) $P_2$ atom, and (f) COHP of $P_1$-$P_2$ interaction in $Cmmm$ $La_2P_3$ at 50 GPa.

To get insight into the formation mechanism of the novel P motifs, we analyze their bonding patterns. Based on the projected electronic density of states (PDOS) of $P6/mmm$ $LaP_2$ (Fig. 2a), there appears a large overlap between La 5d and P 3p orbitals, indicating the formation of ionic bond and a charge transfer from La to P atoms, which is in agreement with Bader charge analysis (Fig. S6). On the other hand, P atom in graphene-like layer has a typical $sp^2$ hybridization in view of a significant overlap between the P 3s, P $3p_x$, and P $3p_y$ states in the energy range from -5 to -15 eV (Fig. 2b), which is further verified from the evident bonding interaction (Fig. 2c). It is well-known that $sp^3$ hybridized P atoms are in favor of forming wrinkled configuration, in which lone pair electrons point in different directions to minimize their repulsive interaction (Figs. 1e-f) [26]. For $LaP_2$, the lone electron pairs are distributed symmetrically on the two sides of a P atom (Fig. 1b). Therefore, the interaction between lone pair electrons and bonding electrons of P atoms in such configuration is balanced, facilitating the formation of a plane. Finally, its stabilization mechanism is distinct from that of graphene-like P/B layer in $LaP_2H_2$ and $MgB_2$. Besides in-plane $sp^2$ hybridized $\sigma$ bond, they also results from the out-of-plane hydrogen bond [28], and π bond [20], respectively.

For $La_2P_3$, the three-coordinated $P_1$ atom also demonstrates an $sp^2$ hybridized feature in view of strong overlap between 3s, $3p_x$, and $3p_y$ orbitals (Fig. 2d), and the linear two-coordinated $P_2$ atom has an sp hybridized character due to the overlap of 3s and $3p_y$ states (Fig. 2e) and obvious bonding interaction (Fig. 2f), which is similar to the bonding pattern of the linear P chain in $Mo_2P$ [51]. To be noted, the lone pair electrons are also symmetrically distributed (Fig. 1d) in $La_2P_3$.

Subsequently, we concentrate on the electronic properties of $LaP_2$. It shows metallization, with the PDOS at the Fermi level ($E_F$) coming mainly from the contribution of La 5d/4f and P 3p orbitals (Fig. 2a). Additionally, there appears the steep bands along the K-Γ and Γ-M directions and the flat bands around the Γ point and H-K direction near the $E_F$ (Fig. 3a), which correspond to the high electron velocity and the large electronic DOS, respectively [52].

Fermi surfaces reflect the electron occupation at the $E_F$. Here, the two Fermi surfaces associated with the bands that contribute significantly at the $E_F$ are explored (Figs. 3b and S7c). One of the Fermi surfaces consists of a capsule-shaped and two tooth-shaped sheets (Fig. 3b), and the other one has porous cylindrical feature (Fig. S7c). The first one is mainly derived from a hybridized state of La 5d ($d_{xy}$ and $d_{z^2}$)/4f ($f_{xz^2}$, $f_{yz^2}$ and $f_{x(x^2-3y^2)}$) and P 3p ($p_x$ and $p_z$) orbitals, and the second one comes from a mixed state of La 5d ($d_{xz}$, $d_{yz}$ and $d_{z^2}$)/4f ($f_{xyz}$ and $f_{yz^2}$) and P 3p ($p_x$, $p_y$, and $p_z$) orbitals (Fig. S9). On the other hand, the former has an obvious nesting along the Γ-A direction (Fig. 3b), which could enhance electron-phonon coupling (EPC) [53,54].

The unique structural and electronic properties of $LaP_2$ inspire us to study its superconductivity. The calculated EPC parameter ($\lambda$) of $LaP_2$ is 1.20 at 16 GPa (Fig. 3c), which is comparable to 1.01 for $Li_6P$ at 270 GPa) [55], 1.19 for $MgP_3$ at 50 GPa [35], and 1.13 for $H_2P$ at 200 GPa [56]. The low-frequency vibrations from the heavy La atoms (below 4.5 THz) contribute 37% to the total $\lambda$, whereas high-frequency vibrations associated with P atoms contribute 63%. Based on the phonon dispersion curves with $\lambda$ weight, the EPC becomes stronger in the Γ point and the Γ-A direction (Fig. 3d), which is associated with the P atom. The former is attributed to the flat-band feature at the Γ point causing a large number of electronic states (Fig. 3a), and the latter is related to the Fermi surface nesting in the Γ-A direction (Fig. 3b). Therefore, we could conclude that the superconductivity of $LaP_2$ mainly originates from the coupling between the La 5d/4f and P 3p orbital electrons and the graphene-like P-derived phonons. Based on the Allen-Dynes modified McMillan equation, the $T_c$ of $LaP_2$ is calculated to be 19.9 K at 16 GPa with a typical Coulomb pseudopotential parameter of $\mu^* = 0.1$.

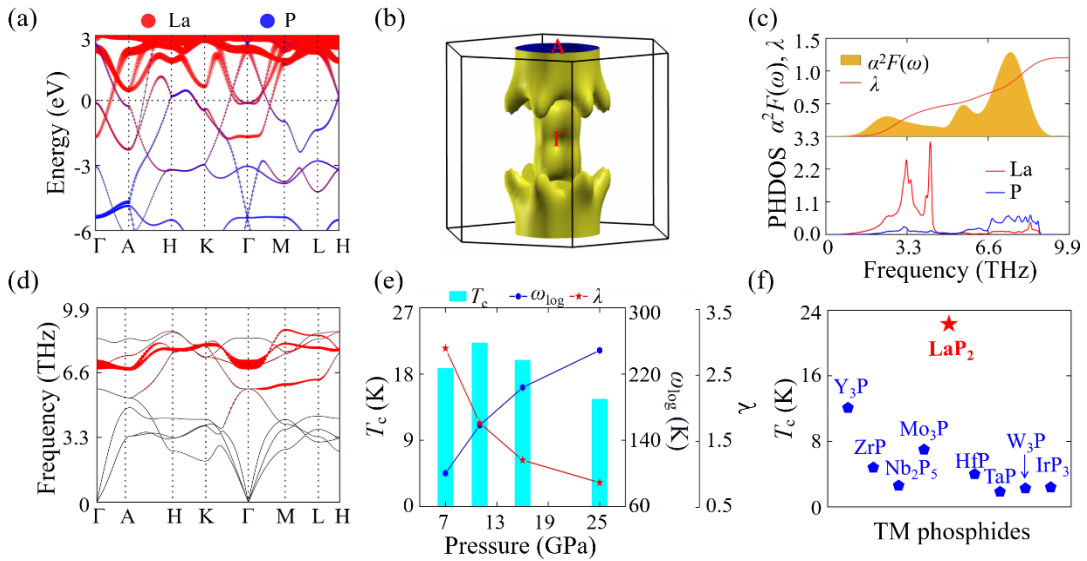

**Fig. 3**. (a) The projected electronic band structure, (b) the nested Fermi surface along the Γ-A direction, (c) the projected phonon density of states (PHDOS), Eliashberg spectral function and frequency-dependent EPC parameters $\lambda$, (d) the calculated phonon dispersion curves at 16 GPa (the magnitude of $\lambda$ indicated by the thickness of the red curves), and (e) pressure-dependent $T_c$ of $P6/mmm$ $LaP_2$. (f) The compared $T_c$ values between $P6/mmm$ $LaP_2$ and TM phosphides [33,57-62].

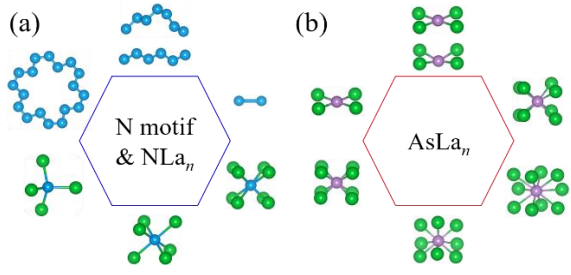

**Fig. 4**. (a) The motifs of polynitrogen and coordination configuration of isolated N atom in La-N compounds, and (b) the coordination configurations of isolated As atom in La-As compounds. Here, $NLa_n$/$AsLa_n$ stands for N-centered/As-centered polyhedral.

$MgB_2$ is a typical BCS superconductor consisting of a honeycomb B layer, and isostructural to $LaP_2$. It is interesting to explore the differences in electronic property and superconducting origin, caused by different atoms in the same configuration. For metallicity, the former comes from the B 2p orbital, and the latter is La 5d/4f and P 3p states. For Fermi surfaces, $MgB_2$ shows cylindrical and horn-like features (Fig. S8) [20], whereas $LaP_2$ is in the form of capsule-shaped, tooth-shaped, and porous cylindrical characters. Although their high-frequency vibrations result from graphene-like B/P layer, the superconductivity of $MgB_2$ is dominated by the electron-phonon coupling of B atoms [20]. For $LaP_2$, La 5d/4f orbital electrons also take part in the coupling. This comparison indicates a possibility to explore distinct superconducting origins in the classical prototype structures (e.g., $H_3S$ and $LaH_{10}$), and broadens the understanding of superconductivity.

For pressure-induced compounds, their lower dynamically stabilized pressure is a key factor for application [63]. The dynamic stability of $LaP_2$ can be down to 7 GPa, which is much lower than $LaBH_8$ [11], a promising low-pressure superconductor. Therefore, we explore its pressure-dependent $T_c$. It is well-known that $\omega_{log}$ and $\lambda$ are the two factors of determining $T_c$. For $LaP_2$, its $\omega_{log}$ increases and $\lambda$ decreases with pressure (Fig. 3e). The change in $\omega_{log}$ can be understood by the fact that the highest phonon frequency gradually raises with pressure, which causes the phonon mode stiffing and $\lambda$ decreasing. Here, $\omega_{log}$ dominates the evolution of $T_c$ below 11 GPa, whereas $\lambda$ becomes a major factor above 11 GPa (Fig. 3e). Thus, 11 GPa is a critical pressure, where shows an optimized $T_c$ of 22.2 K. This might be attributed to the $\lambda$ saturated effect at lower pressure, as found in $CaYH_{12}$ [64]. Additionally, $LaP_2$ has the highest $T_c$ among TM phosphides (Fig. 3f), which is an important type of superconductors [61,65]. Its superconducting transition pressure is also much lower than MoP at 30 GPa [51], TaP at 30 GPa [61], and $IrP_3$ 47.6 GPa [62].

In view of the appearance of P $sp^2$ hybridized planar motif, and N and As belong to the same main-group with P, which encourages us to explore the N/As structural units through their reaction with La at high pressure. Then, we identify several new phases (e.g., $Cc$ $LaN_8$, $R$-$3c$ $LaN_6$, $C2/c$ $LaN_3$, $C2/c$ $LaN_4$, $P$-$3m$1 $La_2N$, as well as $Pm$-$3m$ $LaAs_3$, $Im$-$3$ $LaAs_3$, $P4/nmm$ $LaAs$, $Cmcm$ $La_2As$, Fig. S1), in which N atoms are in the form of N rings, infinite chains, dimers, and isolated atoms, and As atoms prefer to be isolated (e.g. As-centered polyhedra, Fig. 4), whose detailed structures and electronic properties can be found in Figs. S10-15. Some La-N/P/As metallic phases are superconductive, but have the lower $T_c$ values in comparison with $LaP_2$ (Tab. S1).

### IV. CONCLUSIONS

In summary, we have explored the high-pressure phase stability of the La-N/P/As systems through the first-principles structure search calculations. The stable high-pressure stoichiometries and motifs of N, P, and As are different, indicating their distinct high-pressure physical and chemical attributes. In La-P system, $LaP_2$ and $La_2P_3$ exhibit intriguing structural features like a honeycomb P and a planar P layer composed of $P_8$ rings, having $sp^2$ hybridization as well as a mixture of sp and $sp^2$ hybridization, respectively. The superconducting origin of $LaP_2$ is different from the isostructural $MgB_2$. Compared with the reported TM phosphides, $LaP_2$ shows a higher $T_c$ with a lower superconducting transition pressure. Our work not only enriches the P motifs and bonding patterns but also provides impetus to explore superconductivity in similar binary compounds.

**Supplemental Material**
The computational details and phase stability, crystal structures, PDOS, superconducting temperature and structural information of the predicted La-N/P/As compounds, the ELF maps of $La_0P_2$ and $La_0P_3$, and the Fermi surfaces of $LaP_2$ and $MgB_2$.

**Author Contributions**
†X.L. and X.Z. contributed equally.
**Notes**

The authors declare no competing financial interest.
**\*Corresponding Authors**
\*E-mail: <yanggc468@nenu.edu.cn;yanggc@ysu.edu.cn>
**ORCID**
Xing Li: 0000-0002-0250-5382
Xiaohua Zhang: 0000-0002-0434-2863
Guochun Yang: 0000-0003-3083-472X
**Acknowledgments**
The authors acknowledge funding from the Natural Science Foundation of China under 21873017, 52022089, and 21573037, the Natural Science Foundation of Hebei Province A2019203507 and B2021203030, the Postdoctoral Science Foundation of China under grant 2013M541283, and the Natural Science Foundation of Jilin Province (20190201231JC). This work way carried out at National Supercomputer Center in Tianjin, and the calculations were performed on TianHe-1 (A).